\documentclass[final,1p,times]{elsarticle}

\usepackage{multirow}
\usepackage{amssymb}
\usepackage{color}
\usepackage{graphicx}
\usepackage{dcolumn}
\usepackage{bm}
\usepackage{array}
\usepackage{subfig}

\journal{Physics A}

\begin{document}

\begin{frontmatter}

\title{Whether Information Network Supplements Friendship Network}

\author[inst1]{Lili Miao}
\author[inst1,inst2]{Qian-Ming Zhang}
\author[inst1]{Da-Cheng Nie}
\author[inst1]{Shi-Min Cai \corref{cor1}}
\cortext[cor1]{Correspondence author: shimin.cai81@gmail.com}

\address[inst1]{Web Sciences Center, School of Computer Science and Engineering, University of Electronic Science and Technology of China, Chengdu, 611731, P. R. China}
\address[inst2]{Center for Polymer Studies and Department of Physics, Boston University, Boston, 02215, USA,}

\begin{abstract}
  Homophily is a significant mechanism for link prediction in complex network, of which principle describes that people with similar profiles or experiences tend to tie with each other. In a multi-relationship network, friendship among people has been utilized to reinforce similarity of taste for recommendation system whose basic idea is similar to homophily, yet how the taste inversely affects friendship prediction is little discussed. This paper contributes to address the issue by analyzing two benchmark datasets both including user's behavioral information of taste and friendship based on the principle of homophily. It can be found that the creation of friendship tightly associates with personal taste. Especially, the behavioral information of taste involving with popular objects is much more effective to improve the performance of friendship prediction. However, this result seems to be contradictory to the finding in [Q.M. Zhang, et al., PLoS ONE 8(2013)e62624] that the behavior information of taste involving with popular objects is redundant in recommendation system. We thus discuss this inconformity to comprehensively understand the correlation between them.
\end{abstract}


\end{frontmatter}

\section{Introduction}
Connections among individuals aren't created blindly but able to be predicted.
The reference factors could be found in the properties of individuals and their relationships.
Predicting these connections is usually defined as the link prediction problem by treating
individuals and connections as nodes and links respectively\cite{LP2003, LP2011}.
A large number of methods have been proposed to gain more understanding of link creation \cite{LP2011}.
For example, rich-get-richer mechanism suggests that nodes with more
neighbors will attract new links with higher probability \cite{BA1999},
clustering mechanism declares that two nodes have a high probability of
creating a link between them if they share some common neighbors \cite{Cluster},
and homophily mechanism states the observed tendency that people tie with others
of similar profiles or experiences \cite{Homophily}. All these mechanism and
their derivation are widely used in link prediction problem.

Here we mainly concern on homophily mechanism that is confirmed by plenty of evidences,
such as the observed tendency
in an acquaintance network of university members \cite{HomoExam1},
a large-scale instant-messaging network containing $1.8\times10^8$ individuals \cite{HomoExam2},
friendship networks of a set of American high schools \cite{HomoExam3},
a social network of a cohort of college students in Facebook \cite{HomoExam4},
and performing well in predicting missing links for directed networks co-working with potential theory \cite{QMpotential}. Homophily could also be reflected
in individual reading and musical tastes, shopping history, interests,
and so on. These information is widely used to measure the similarity
between users in collaborative filtering recommender systems \cite{RS2012}.
Meanwhile, facilitated by online social networks (e.g., \emph{www.epinions.com} and \emph{www.last.fm}),
users can not only create connections to others, but also choose objects they like.
From the perspective of homophily, two users can be considered similar if they
have many common friends, or have chosen plenty of same objects.
Here we raise the question: are the two kinds of similarity correlated with each other?

Some researchers have already revealed that the preferences
of trusted friends were beneficial to improve the effectiveness of
recommendation \cite{trust1, trust2, trust3}. In turn, it is lacking
that whether the similarity of personal taste could reflect the intimate friendships.
This question is pregnant as the friendship and personal taste commonly
characterize the users belonging to an identifiable social network.
It is beneficial to check the consistency between the two features of homophily,
denoting the relationship type and personal taste. To address this question,
we collect two data sets both including  users' friendships and behavior information,
i.e., reading traces in \emph{epinions.com} and listening traces in \emph{last.fm}.
The similarities between pairs of users are measured according to their friendships and personal tastes.
Experimental results suggest that the similarity of taste is helpful but doesn't work solely.
Further, considering the possible redundant or misleading information in recommender systems \cite{RSbackbone},
we reexamine the effect of similarity of taste through gradually removing the behavior information by different strategies.
It interests us that these popular objects, which were considered to reflect common
taste and lower personalized information in Ref. \cite{RSbackbone}, play much more
important role in improving the effective of link prediction in social network.

The rest of paper is organized as follows: In section 2 we introduces the data sets and link prediction model, and
in section 3 present the experimental results and the discussion. Finally, we show our conclusion in section 4.

\section{Material and Method}

\subsection{Data}
We find two available data sets that consist of both preference of users and relationship among them.
One is collected from \emph{Lastfm} \footnote{www.last.fm}, an online music system
where each user has a list of most listened songs and is allowed to make friends
with others online \cite{lastfm}. The other is gathered from \emph{Epinions} \footnote{www.epinions.com},
a general consumer review site where users can read reviews about variety of items
and allows users to decide whether to trust others \cite{epinions}. To represent these data clearly,
we build two networks for each data set: Relation Network (RN) and Taste Network (TN). Concretely,
RN indicates the relationship between users, while TN represents the preference of users to objects.
We name them as LastfmRN, LastfmTN, EpinionRN and EpinionTN, respectively.
\begin{itemize}
  \item LastfmRN contains 1543 online users interconnected by 25434 connections.
  \item LastfmTN is a bipartite network which contains 1543 users and 18745 songs. There are 92834 links between users and songs in total.
  \item EpinionRN consists of 49288 users and 381035 trust relations.
  \item EpinionTN is a bipartite network with 664823 ratings on 139738 reviews rated by 49288 users.
\end{itemize}

\subsection{Similarity-based Link Prediction Model}
An abundant of indices have been proposed to measure the node similarity in networks \cite{LP2011}.
For effectiveness and simplification, we choose resource allocation (RA) index \cite{RS-RA,LP-RA}
identified as a well performed one based on local information. Considering a pair of nodes, $x$ and $y$, which aren't
directly connected. The node $x$ can send some resource to $y$, with their common neighbors playing the role of transmitters.
In the simplest case, we assume that each transmitter has a unit of resource, and will averagely distribute to the all neighbors.
Then, the similarity between $x$ and $y$ can be defined as the amount of resource $y$ received from $x$,
\begin{equation}
  s_{xy}=\sum_{z\in\Gamma(x)\cap\Gamma(y)}\frac{1}{k(z)} ,
  \label{eq-RA}
\end{equation}
where $\Gamma(x)$ and $\Gamma(y)$ are the set of neighbors of $x$ and $y$ respectively and
$k(z)$ is the degree of node $z$. Note that the common neighbor $z$ has different meanings
in RN and TN. In RN, $z$ indicates common users; while in TN, the common neighbors of
users are obviously the objects.

The similarity-based link prediction model assumes that the higher similarity a pair of disjointed nodes have, the more probable a link would exist between them.
We employ the AUC (area under the receiver operating characteristic curve) to evaluate the performance of link prediction index. AUC \cite{AUC1983} can be interpreted as the probability that a randomly chosen missing link
(a link in $E^P$) is given a higher score than a randomly chosen nonexistent link (a link in $U\backslash E$, where $E$ is the set of links in the given network, $U$ denotes the universal set of links). To implement AUC, the links of RN $E$
need to be randomly divided into two parts, the training set $E^T$ with $90\%$ links is treated as known information
while the testing set (probe set) $E^P$ with $10\%$ links is used for testing.
Clearly, $E=E^T\bigcup E^P$ and $E^T\bigcap E^P={\O}$. Among $n$ independent comparisons, if there are $n'$ times the missing link having
a higher score and $n''$ times they have the same score, then we define AUC as $\mathrm{AUC}=(n'+0.5n'')/n$.
Suppose all the scores are generated from an independent and identical distribution, the AUC should be about $0.5$.
Therefore, the degree to which the value exceeds $0.5$ indicate a better performs than random prediction.

\section{Experiments}

\begin{figure*}[ht]
\centering
  \includegraphics[width=5cm,height=3.8cm]{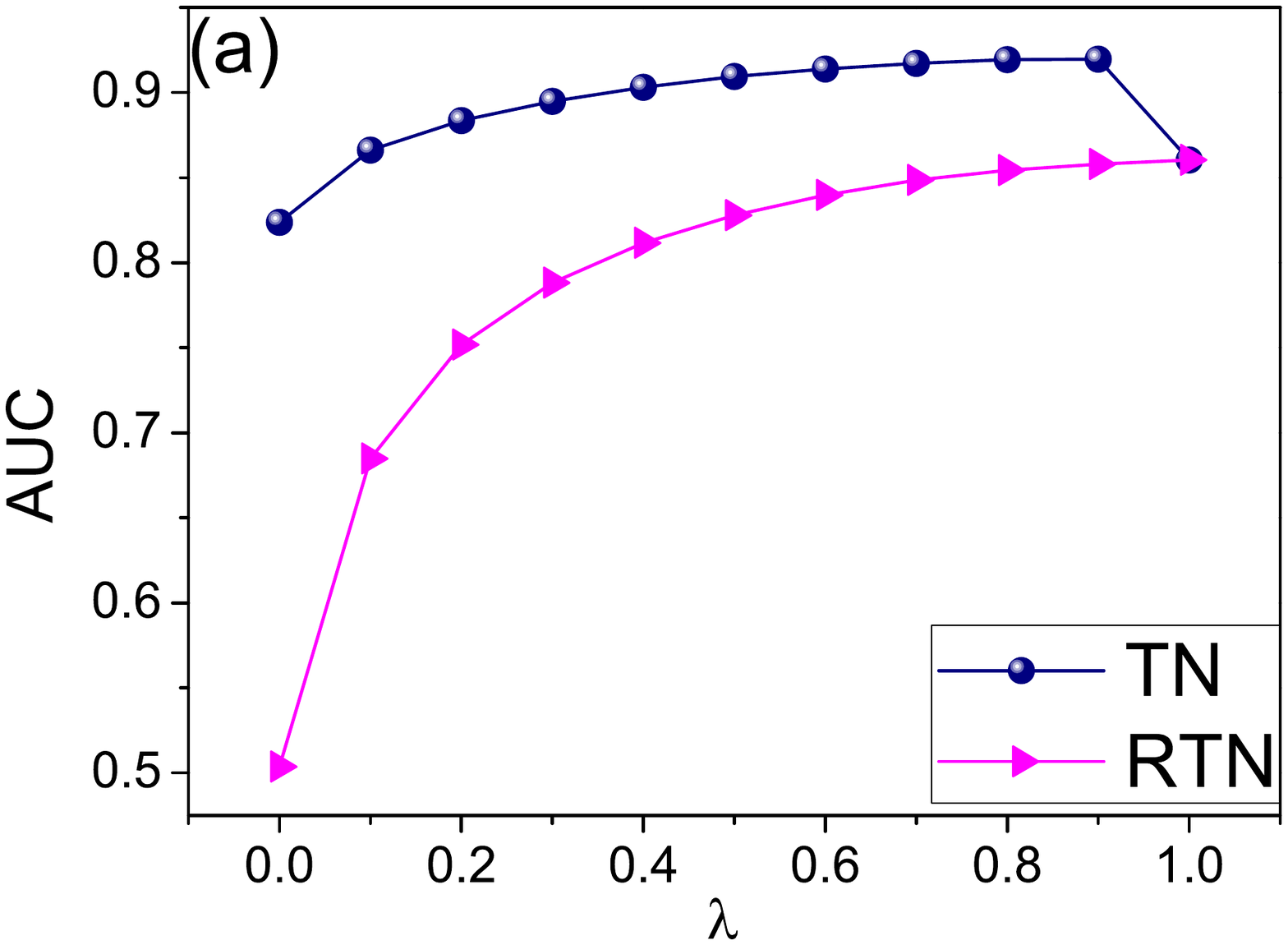}
  \includegraphics[width=5cm,height=3.8cm]{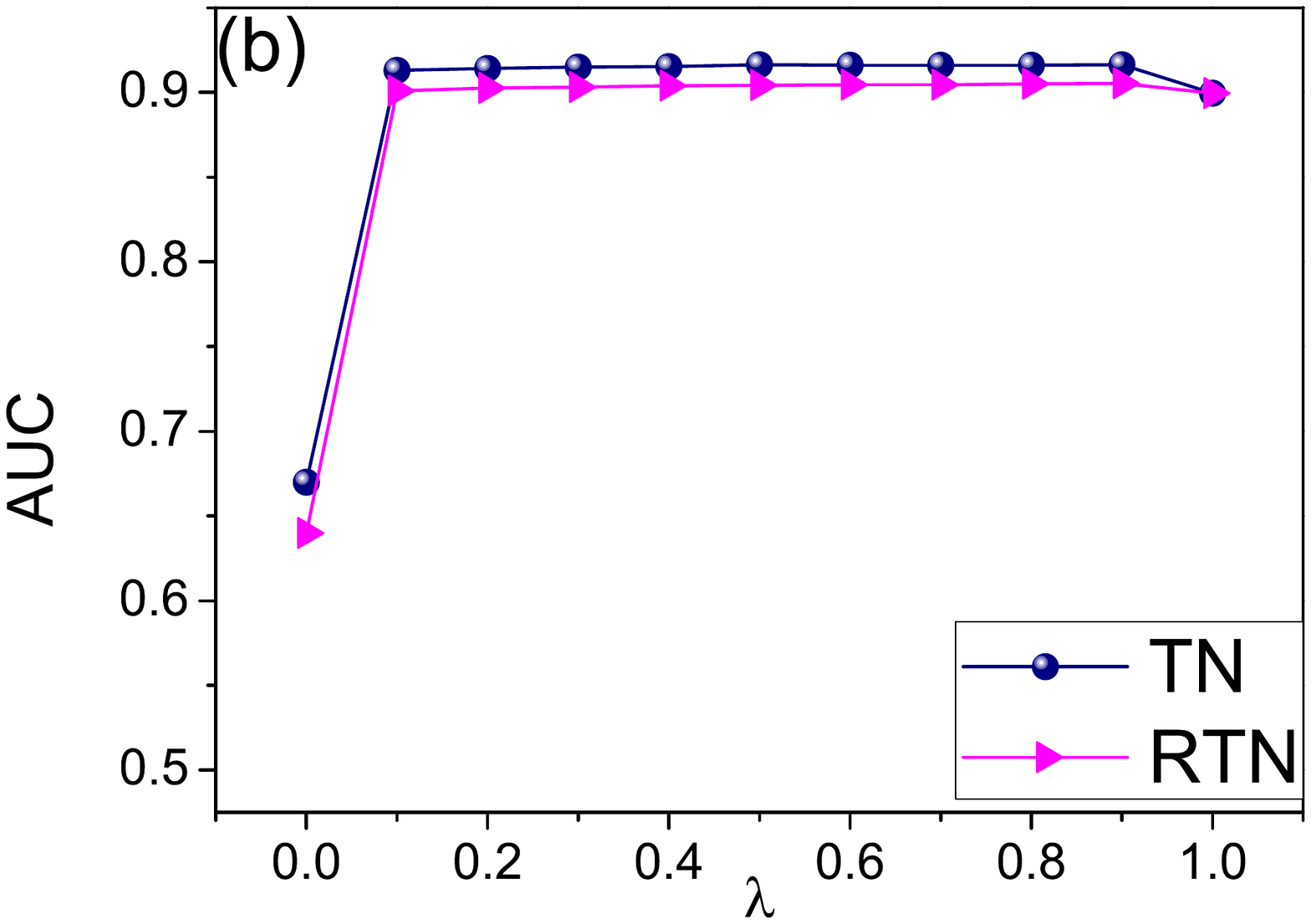}
  \caption{(Color online) The performance of predicting missing friendships by introducing the similarities of personal tastes.
  Each subfigure corresponds to result in different data sets (a) Lastfm and (b) Epinions. The purple curve indicates
  the result brought by the null model, which obviously provide a strong evidence to
  support that personal taste is related to the formation of friendship.}\label{fig-shuffle}
\end{figure*}

A direct way of measuring the effect of similarity of taste between two users is that apply it to predict missing friendship by
introducing a tunable parameter $\lambda$. According to Eq. (\ref{eq-RA}), a hybrid predictor is defined as
\begin{equation}
  s_{xy}=\lambda\cdot s_{xy}^{\mathrm{RN}}+(1-\lambda)\cdot s_{xy}^{\mathrm{TN}},
  \label{eq-hybRA}
\end{equation}
where $s_{xy}^{\mathrm{RN}}$ is the similarity between user $x$ and user $y$ based on RN, while $s_{xy}^{\mathrm{TN}}$
is based on TN. The AUC based on this hybrid predictor with different $\lambda$ are shown in Fig. \ref{fig-shuffle}.
Meanwhile, for comparison, we also represent the AUC brought by randomized TN (the so-called Null Model of networks \cite{milo2002}),
which are produced by shuffling links more than $5\times |E'|$ times to insure the network is random enough
(the $E'$ is the set of links in TNs). It is obvious that simply mixing these two types of similarity can
improve the prediction accuracy effectively; while when the TN are randomized, the information therein is not so helpful any more. The results provide a strong evidence to support that personal taste is related to the formation of friendship. Besides, we can also notice that in Fig. \ref{fig-shuffle}(b) the difference between the results of AUC for TN and RTN aren't obvious, and the peak value for RTN does not fall on $\lambda=1$. To answer them, we need to know how the behavior information affects the measurement of similarity between users. We will provide the explanation in the following part, but firstly we employ some different algorithms to figure out what kind of information can be the best representation on behalf of the individual's behavior, inspired by the information backbone study \cite{RSbackbone}.

Information backbone for recommender systems revealed that some links in the online user-object
bipartite networks were redundant or even misleading \cite{RSbackbone}. In other word, the similarity
between users might be measured not so exactly because of the misleading information.
Then we want to know whether the so-called misleading links would reduce the prediction accuracy.
Thus, we conduct the further study to check the tendency of prediction accuracy by gradually removing links from TN.
Restricted by the data set, we only take the topology-aware algorithms into consideration and compare them with the strategy of random removal:
\begin{enumerate}
  \item [(a)] Most popular removal (MPR): The popularity of a link ($u$, $o$) is defined as $k_u \times k_o$, where $k_u$ ($k_o$) is the degree of user $u$ (item $o$). We calculate the popularity of all the remaining links and remove the most popular links. Need to notice that, the popularity of every link should be renewed after each removal, and so does the popularity for the following methods.
  \item [(b)] Least popular removal (LPR): The most unpopular links will be removed.
  \item [(c)] Most popular removal for each user (MPRE): The most popular link for each target user is removed.
  \item [(d)] Least popular removal for each user (LPRE): The most unpopular link for each target user is removed.
  \item [(e)] Random removal (RR): Link is randomly chosen and removed.
\end{enumerate}
In order to make all the algorithms comparable, all links are removed in $10$ macro-steps. Therefore, about $10\%$ links
will be chosen in each macro-step. For example, if there are 45 links in the original network, on average $45/10=4.5$ links should
be removed in each macro-step, and after $n$th macro-step, $\lceil4.5n\rceil$ links will be removed from the network.
In MPRE and LPRE algorithms, the number of links to be removed at each marco-step for each user is proportional to his degree.

\begin{figure}[ht]
  \centering
  \includegraphics[width=4cm]{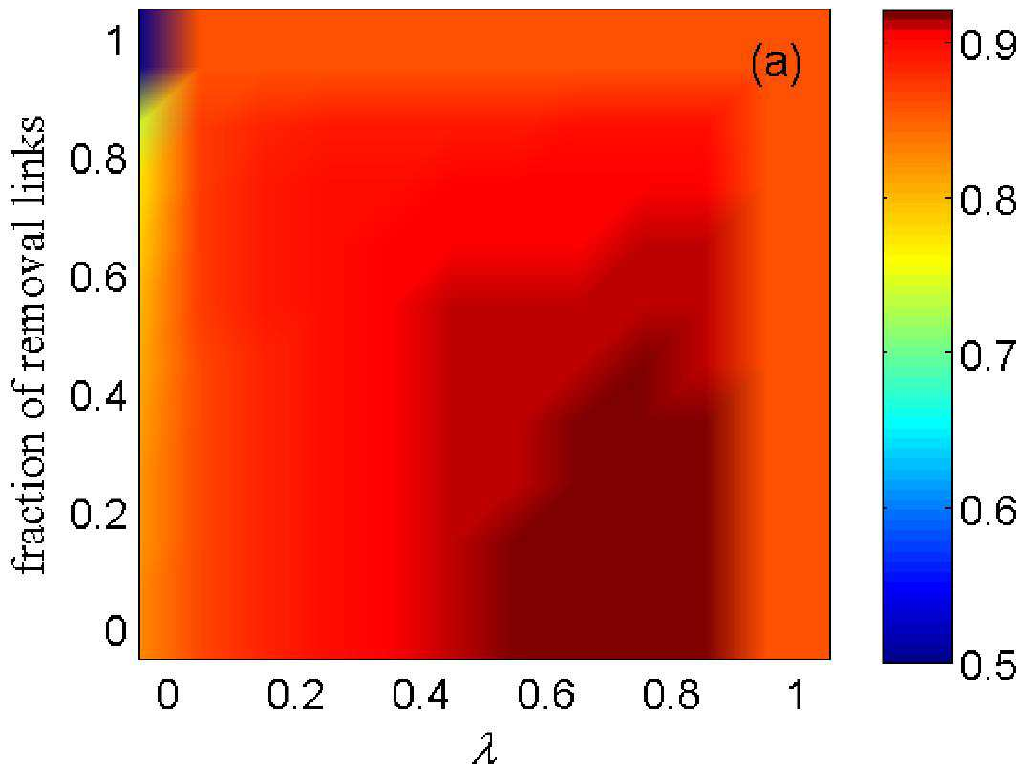}
  \quad
  \includegraphics[width=4cm]{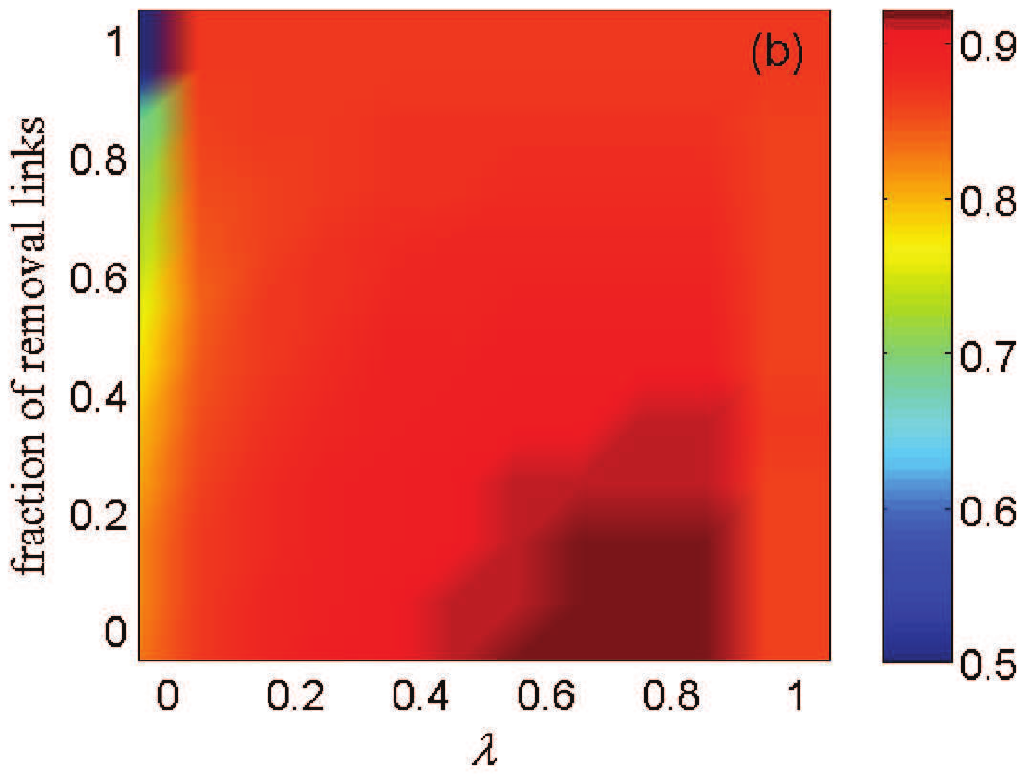}
  \quad
  \includegraphics[width=4cm]{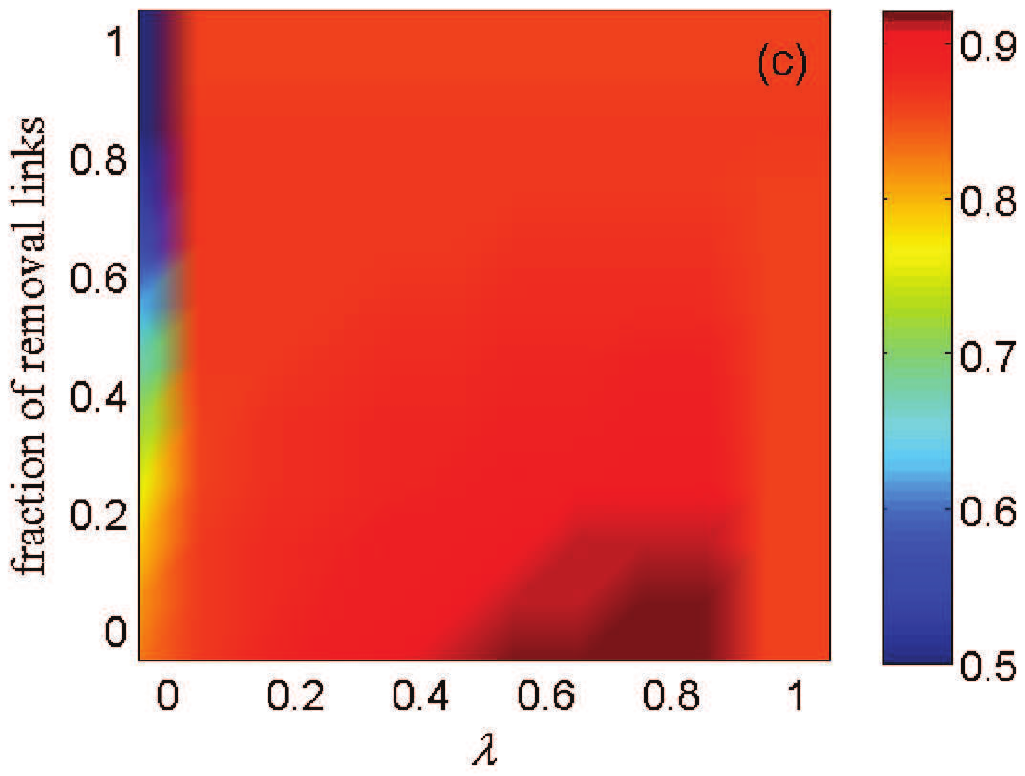} \\
  \quad \\
  \includegraphics[width=4cm]{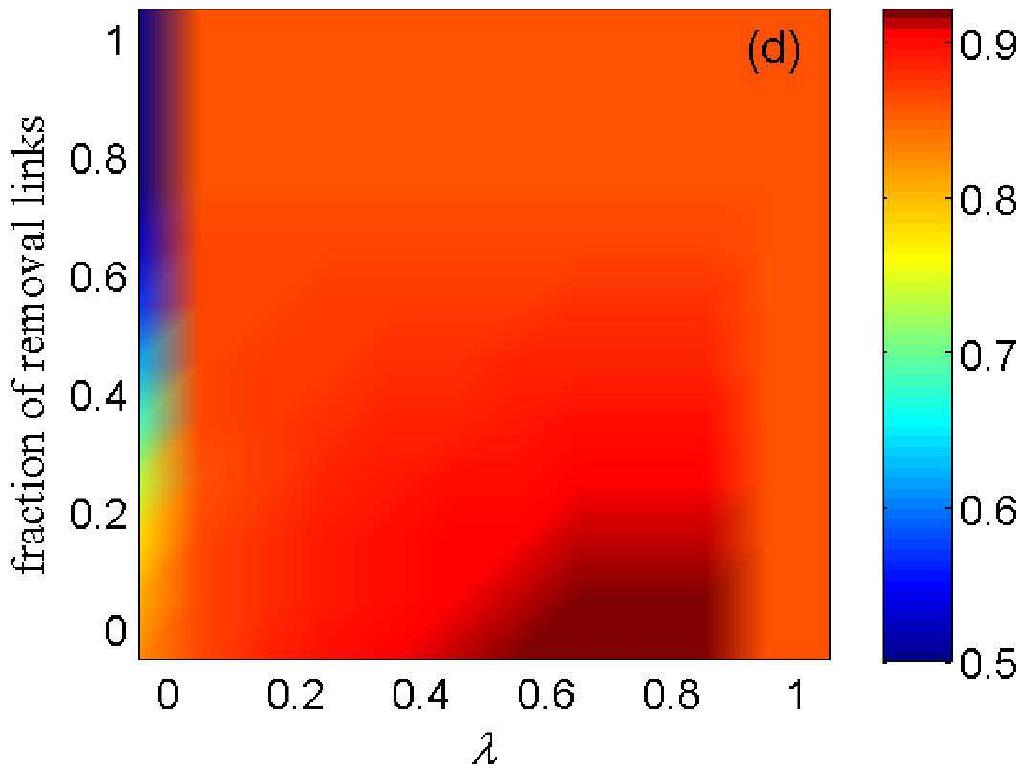}
  \quad
  \includegraphics[width=4cm]{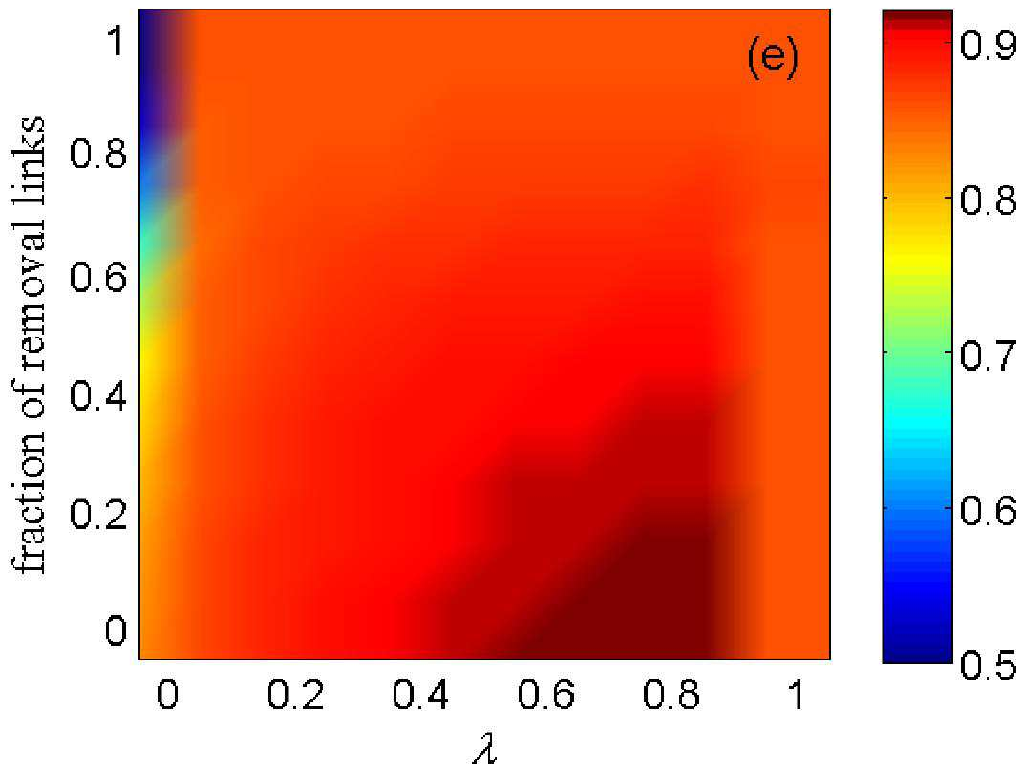}
  \caption{(color online) Comparisons of prediction accuracies among different strategies of removing links in Lastfm.
  The horizontal axis $\lambda$ is to adjust the proportion of the two kinds of similarities, which is defined in Eq. (\ref{eq-hybRA}).
  The vertical axis indicates the fractions of removed links. Each subfigure corresponds to a strategy, (a) LPRE; (b) LPR;
  (c) MPRE; (d) MPR; (e) RR, respectively.}\label{fig-Removal}
\end{figure}

\begin{figure}[ht]
  \centering
  \includegraphics[width=4cm]{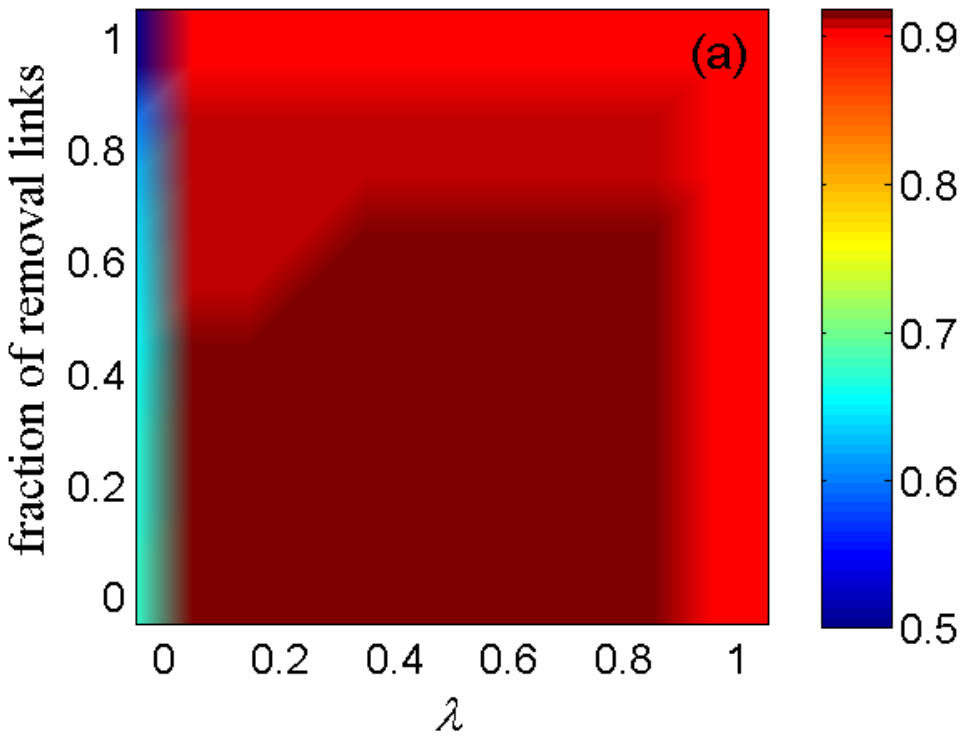}
  \quad
  \includegraphics[width=4cm]{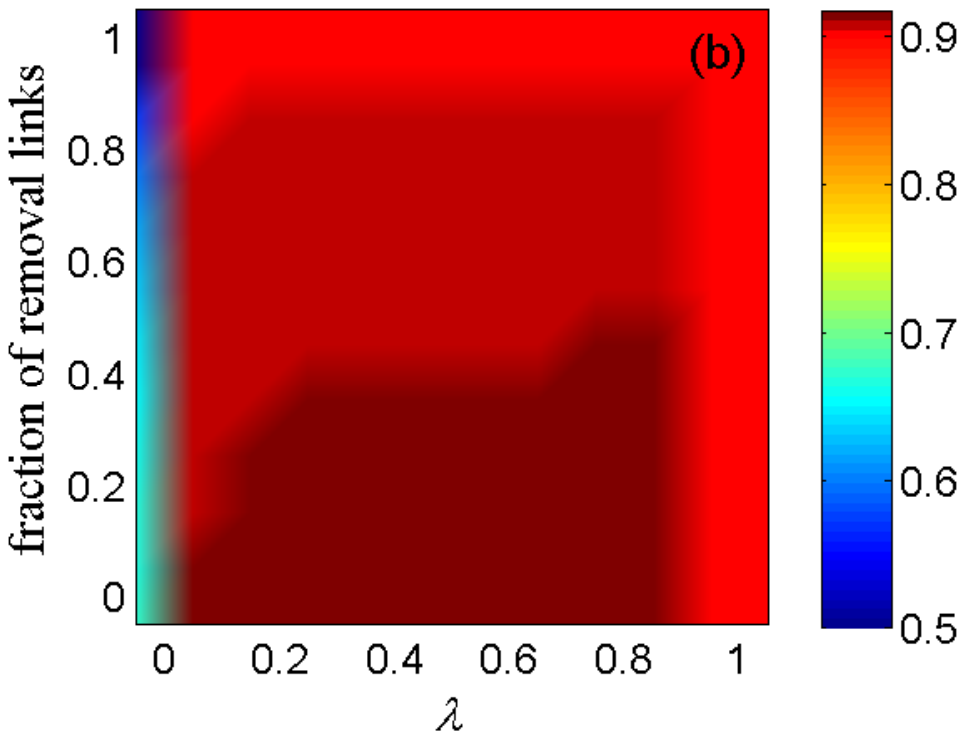}
  \quad
  \includegraphics[width=4cm]{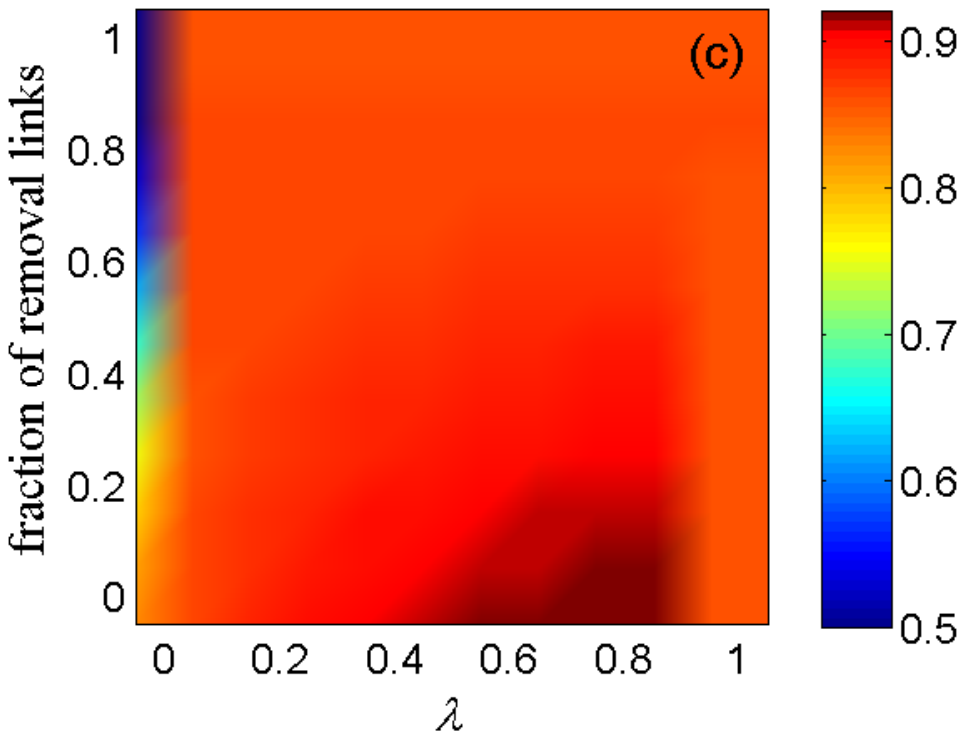} \\
  \quad  \\
  \includegraphics[width=4cm]{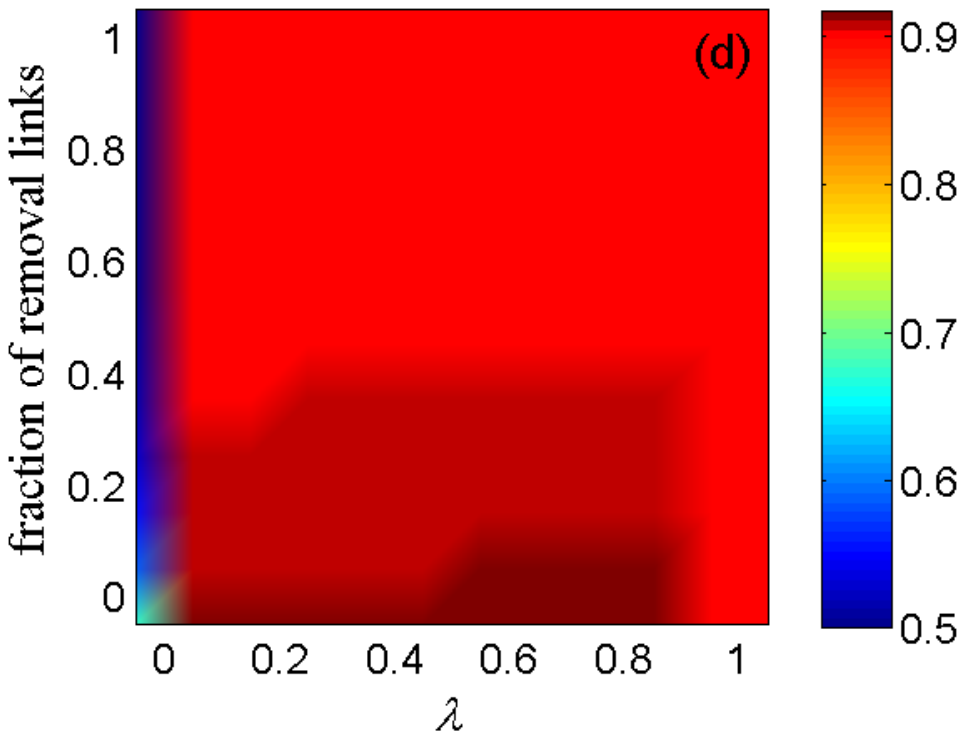}
  \quad
  \includegraphics[width=4cm]{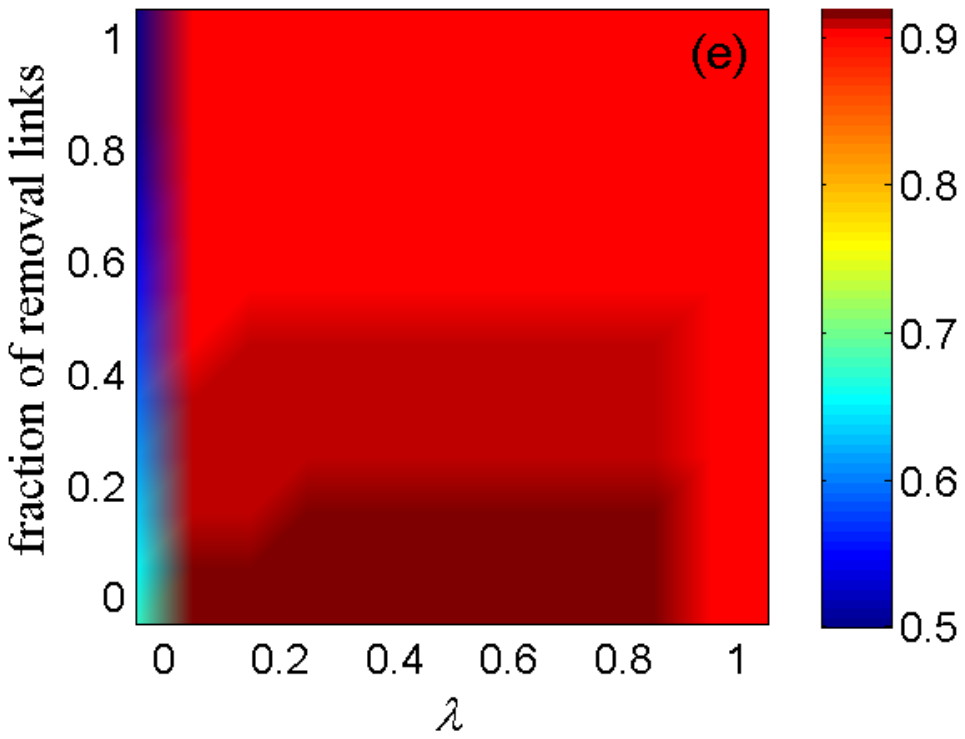}
  \caption{(color online) Comparisons of prediction accuracies among different strategies of removing links in Epinions.
  The horizontal and vertical axis is same as in Fig \ref{fig-Removal}. Each subfigure corresponds to a strategy, (a) LPRE; (b) LPR;
  (c) MPRE; (d) MPR; (e) RR, respectively.}\label{fig-RemovalofE}
\end{figure}


In the simulation, links are gradually removed from TN according to different algorithms. For each macro-step, we study
the change of link prediction performance by combining the similarity of the rest TN. The results of Lastfm and Epinions,
showing a similar change trend for each algorithm are presented in Fig. \ref{fig-Removal} and Fig. \ref{fig-RemovalofE}, respectively. Overall speaking, LPR and LPRE are much better than MPR and MPRE. Especially for LPRE, its AUC can be maintained even after half of the links are removed. However, in the cases of MPR and MPRE, their AUCs begin to decline when removed only $20\%$ links. These results strongly suggest that the links with high popularity in TN are more helpful to predict missing friendships. And whether can we come to the conclusion that the links with high popularity can mostly represent the TN? Nevertheless, Zhang et al. \cite{RSbackbone} have claimed that the links with low popularity generally contain more information, which are able to improve higher accuracy of TN-based recommendation system. These two results seems to be contradictory, urging us to
find an intuitive reasonable explanation.

\begin{figure}
  \centering
  \includegraphics[width=5cm,height=3.8cm]{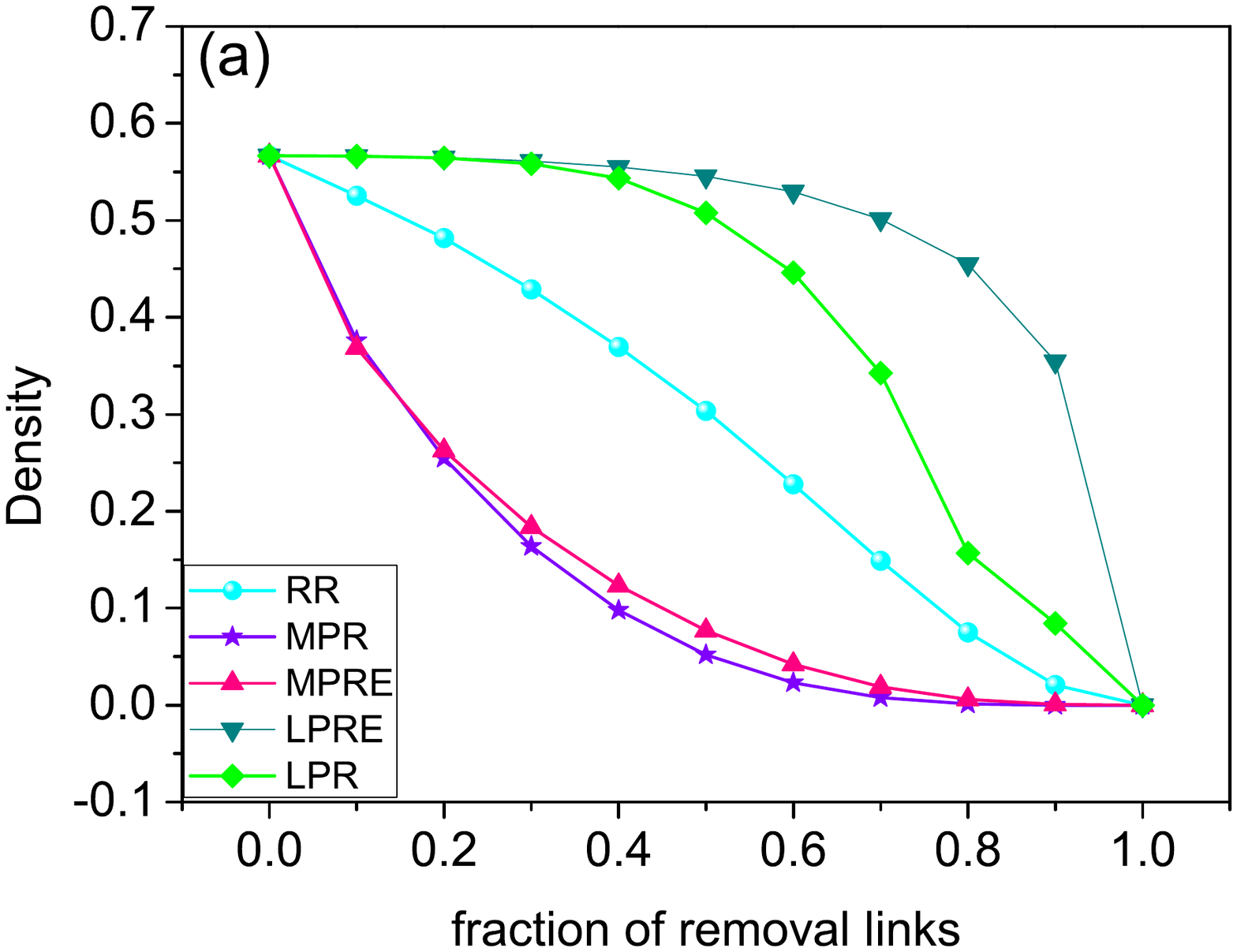}
  \includegraphics[width=5cm,height=3.8cm]{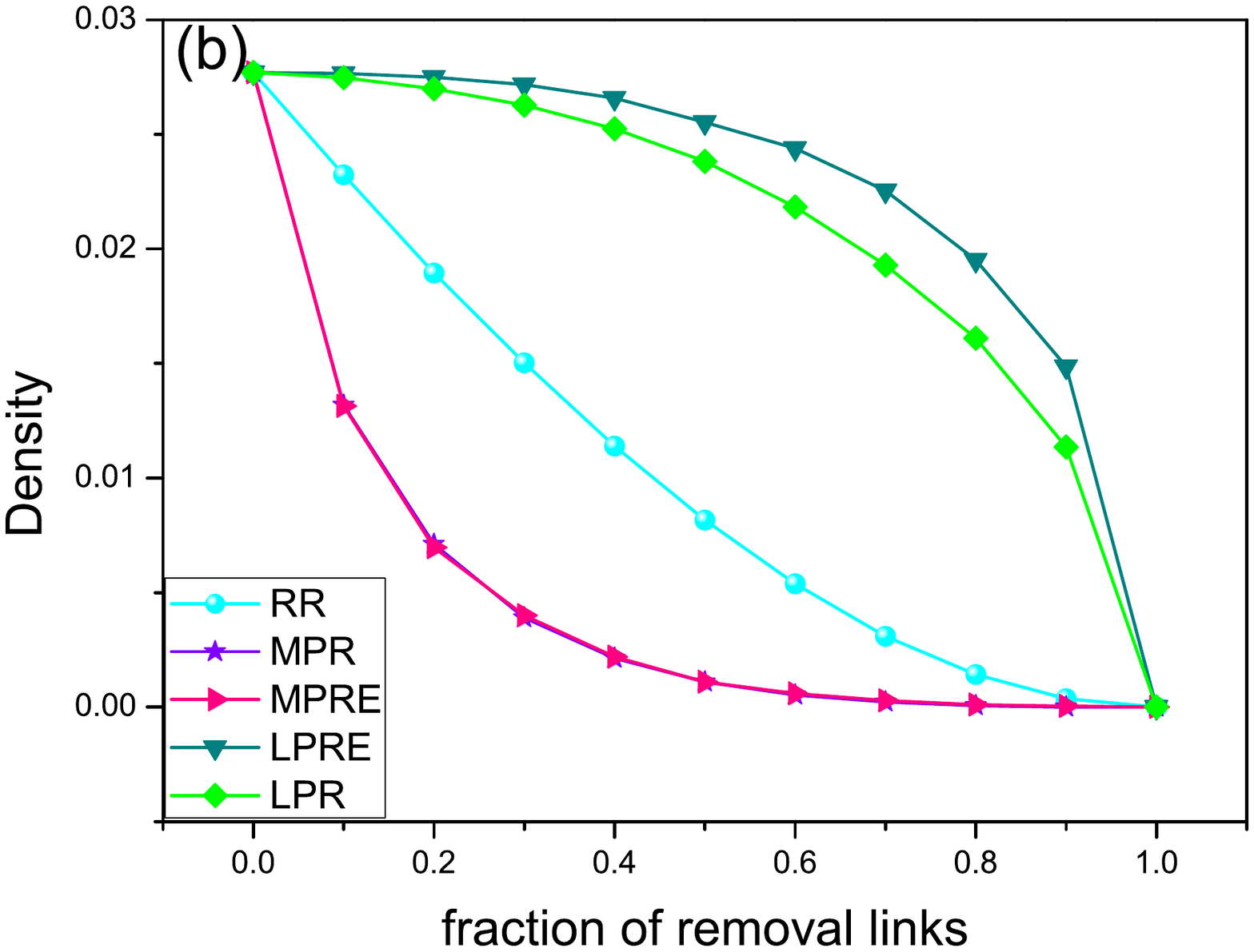}
  \caption{Comparison of the density of the user projection networks after each macro-step between different
  strategies of removing links. Each subfigure corresponds to result in different data sets,
  (a) Lastfm and (b) Epinions.}\label{fig-density}
\end{figure}

Recalling the calculation of similarity in TN, we find the similarity is larger than $0$ if the
two users have at least one common neighbor (i.e., object). Thus, it is essentially the same as the process of bipartite network projection \cite{Projection}. And, the density of link in user projection network of TN where a link (i.e., user-object-user triangular relationship) will be created between two users if they have at least one common neighbor in TN, represents how much information of similarity can be obtained from the TN. For these five strategies of link removal, we build all the user projection networks based TN after each macro step, and compute the density of every projection network. As we expect, the tendency is in
accordance with the performance of the removal algorithms.
As Fig. \ref{fig-density} shows, LPR and LPRE can keep more
information of similarity between users than others.

\begin{figure}
  \centering
  \includegraphics[width=5cm,height=3.8cm]{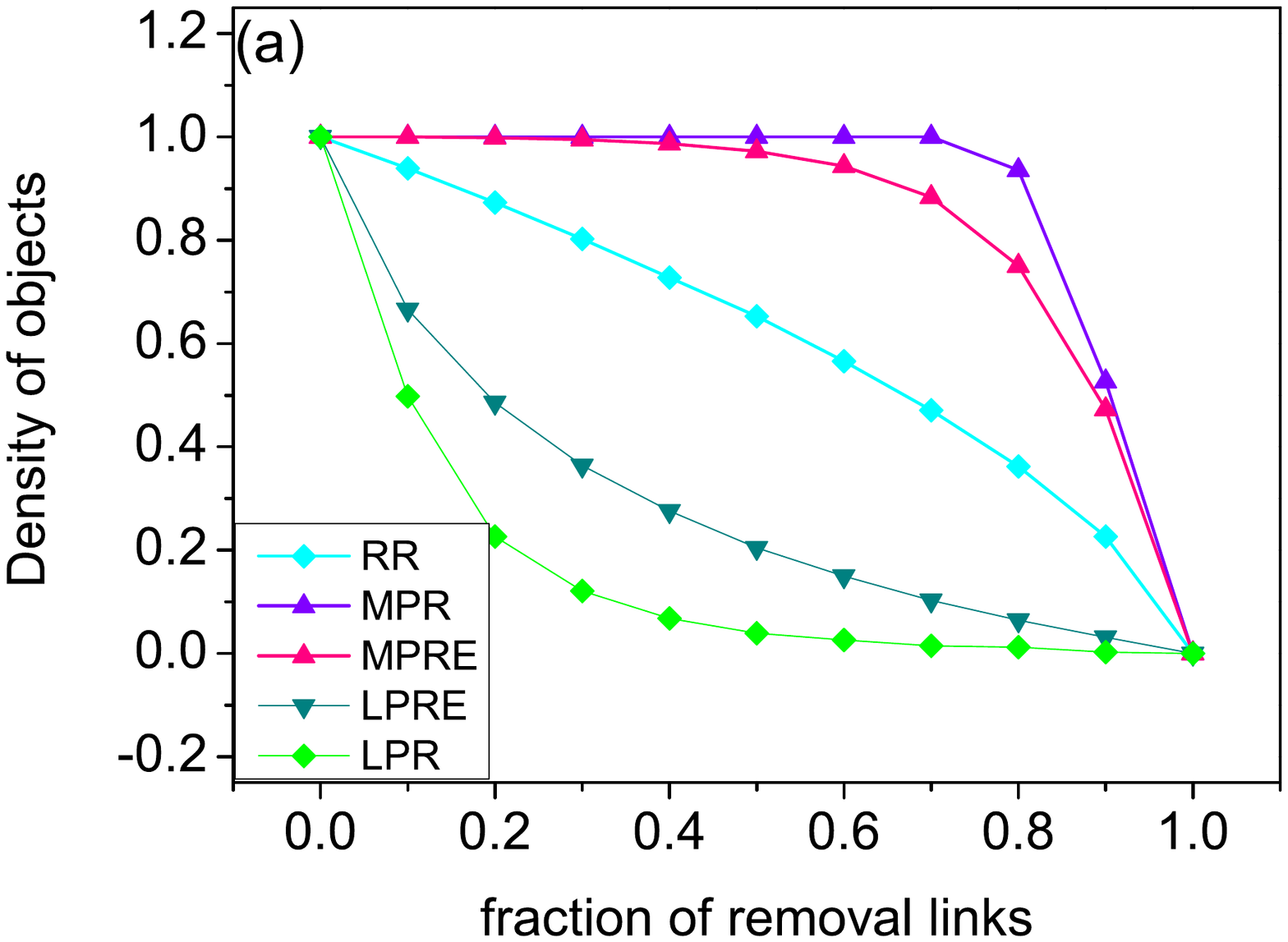}
  \includegraphics[width=5cm,height=3.8cm]{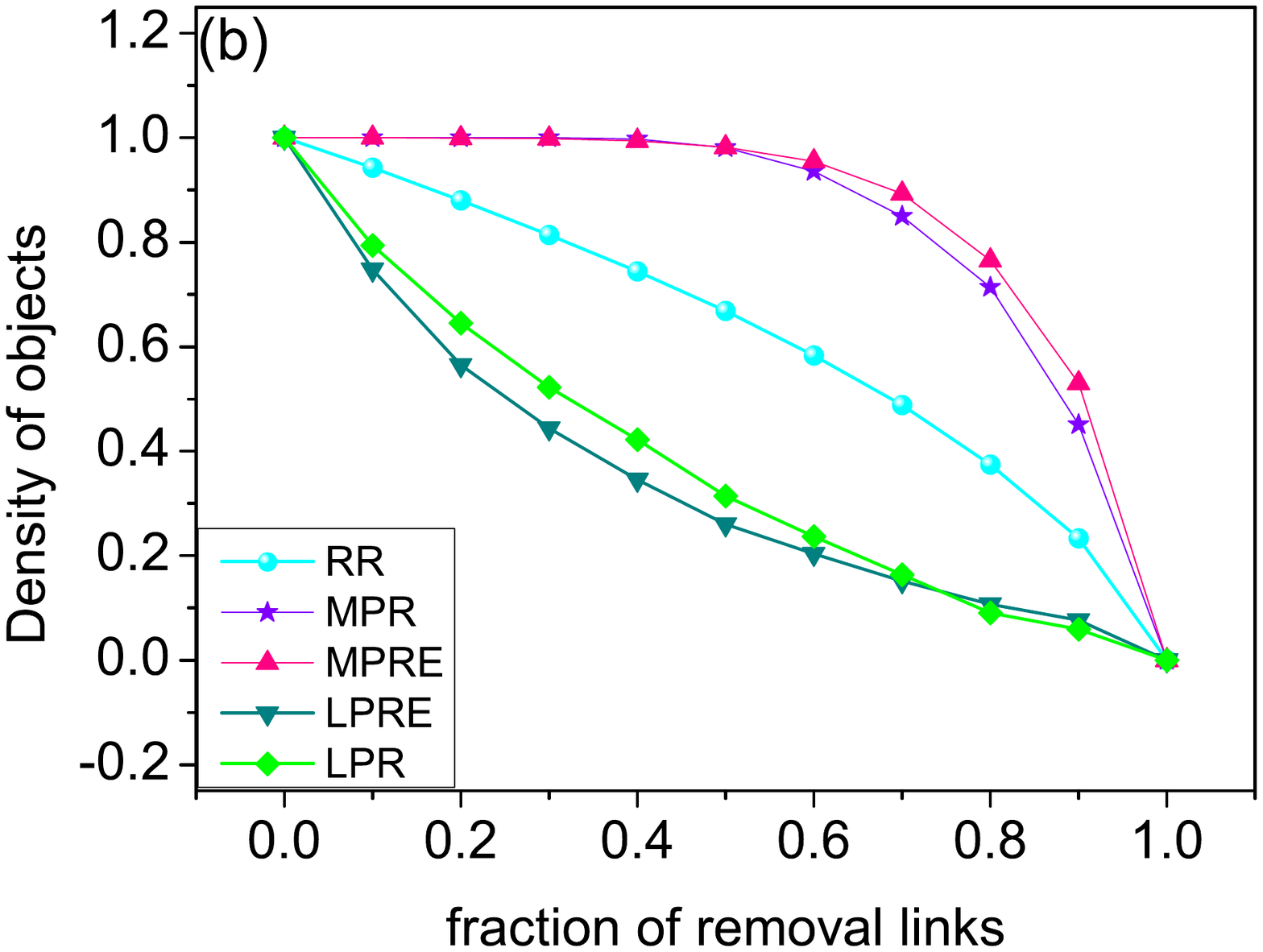}
  \caption{Comparison of the density of objects whose degree are larger than 0 in TN at each macro-step for 4 strategies.
  Each subfigure corresponds to result in different data sets, (a) Lastfm and (b) Epinions.}\label{fig-Objects}
\end{figure}

More deeply, we perform an analysis on the types of usable information obtained from TN in
predicting missing friendships and recommendation system, respectively. In the process of
link removal in TN, at each macro-step, we statistically count the number of objects whose
degree is larger than 0, and scaled it with total objects (i.e., the density of objects).
As shown in Fig. \ref{fig-Objects}, on one hand, LPR and LPRE quickly make unpopular objects (corresponding to
their links with low popularity) be isolated yet remain much more popular ones (corresponding to
their links with high popularity) in TN. These popular objects keep much more
user-object-user triangular relationships indicating the usable information of similarity,
which is consistency with these results in user projection network. On the other hand,
in TN-based recommendation system, recommending process tries to create user-object
links and form more rectangles. The isolated objects resulting from LPR and LPRE
cannot be recommended precisely to users. On the contrary, MPR and MPRE only reduce the redundant information
of popular objects but still maintain as many as objects of recommendation system. Thus,
the usable information obtained from TN for predicting missing friendships
emphasizes user-object-user triangular relationships, while the usable information in
TN-based recommendation system describes user-object relationships. Based on the above-mentioned
analysis,  there is no conflict from the perspective of usable information, but both
aim to keep more useful information for themselves.

Then, we can also explain the difference between the two curves in Fig. \ref{fig-shuffle}(b).
As discussed above, the effectiveness of behavior information depends on the density of the projection networks.
As shown in Fig. \ref{fig-density}, it can be found that the density for Lastfm is mostly up to 0.57, while that for Epinions is even below 0.03.
Thus, most of the similarities between users of Epinions are 0 when considering only the behavior information.
As a result, although we exchange the links in TN, not many pairs of users whose similarity is 0 would newly appear which makes the
difference of AUC in Fig. \ref{fig-shuffle}(b) be not obvious.


\section{Conclusion}
The homophily among people can be measured by many factors, which usually exist simultaneously in one system.
Herein, we consider not only the common friendships among people, but also similar preferences to objects (i.e., tastes).
To examine the consistency between these two dimensional factors, we collected two data sets that include online friendships
and personal behavioral information involving with his/her preference to objects.
Through link prediction analysis, we show that their tastes are very helpful to predict missing friendships, and also find that
the popular objects preferred by users play a much more important role.
We note that these results seem to be be contradictory to the conclusion in Ref. \cite{RSbackbone}, that is, the
popular objects are redundancy in recommendation system. A comprehensive analysis from the perspective of usable information
is presented to explain this inconformity, which suggests that the user-object-user triangular relationships and
user-object ones both obtained from TN are respectively key to friendship prediction and recommendation system.
Moreover, to confirm our results are independent on link prediction method,
we perform extensive experiments and test several mainstream indices, such as Common Neighbor Index, Adamic-Adar Index, Jaccard Index and Salton Index.
All of them show approximate results and suggest the robustness of conclusion.
Finally, we state that the same information may play opposite roles in different problems,
that is the the effectiveness of information depends on the specific issues and how it works
although the information is collected from the same system.


\section{Acknowledgments}
Thanks for the helpful discussion with Tao Zhou. This work is jointly supported by the NNSFC (Grant Nos. 91024026 and 11205042), Special Project of Sichuan Youth Science and Technology Innovation Research Team (No.2013TD0006), and the Fundamental Research Funds for the Central Universities (Grant No and ZYGX2012J075). QMZ acknowledges the support from the Program of Outstanding PhD Candidate in Academic Research by UESTC (No. YBXSZC20131034) and China Scholarship Council.

\end{document}